\documentclass[twocolumn,showpacs,superscriptaddress]{revtex4}

\usepackage{epsf}
\usepackage{epsfig}
\usepackage{graphicx}
\usepackage{dcolumn}
\usepackage{bm}
\usepackage{amsfonts}
\usepackage{amsmath}
\usepackage{amssymb}
\usepackage{color,soul}

\newcommand{\unit}[1]{\ensuremath{\, \mathrm{#1}}}

\input epsf

\begin{document}

\title{Tuning laser-induced bandgaps in graphene}
\author{Hern\'{a}n L. Calvo}
\affiliation{Instituto de F\'{\i}sica Enrique Gaviola (IFEG-CONICET) and
FaMAF, Universidad Nacional de C\'{o}rdoba, Ciudad Universitaria, 5000
C\'{o}rdoba, Argentina.}
\affiliation{Institut f\"{u}r Theorie der Statistischen Physik, RWTH Aachen University, D-52056 Aachen, Germany.}
\author{Horacio M. Pastawski}
\affiliation{Instituto de F\'{\i}sica Enrique Gaviola (IFEG-CONICET) and
FaMAF, Universidad Nacional de C\'{o}rdoba, Ciudad Universitaria, 5000
C\'{o}rdoba, Argentina.}
\author{Stephan Roche}
\affiliation{CIN2(ICN-CSIC) and Universidad Aut\'{o}noma de Barcelona, 
Catalan Institute of Nanotechnology, Campus UAB, 08193 Bellaterra (Barcelona), Spain.}
\affiliation{ICREA, Instituci\'{o} Catalana de Recerca i Estudis Avan\c{c}ats, 08070 Barcelona, Spain.}
\author{Luis E. F. Foa Torres}
\affiliation{Instituto de F\'{\i}sica Enrique Gaviola (IFEG-CONICET) and
FaMAF, Universidad Nacional de C\'{o}rdoba, Ciudad Universitaria, 5000
C\'{o}rdoba, Argentina.}
\date{\today}

\begin{abstract}
Could a laser field lead to the much sought-after tunable bandgaps in
 graphene? By using Floquet theory combined with Green's functions
 techniques, we predict that a laser field in the mid-infrared range can
 produce observable bandgaps in the electronic structure of
 graphene. Furthermore, we show how they can be tuned by using the laser
 polarization. Our results could serve as a guidance to design opto-electronic
 nano-devices. 
\end{abstract}

\maketitle

More than a century ago, the use of alternating currents (ac) sparked a
revolution that changed our modern world. Today, the use of ac fields
has reached the nanoscale \cite{Platero2004,Kohler2005}. Here, the
interplay between the quantum coherence of the electrons, inelastic
effects and dynamical symmetry breaking offers fascinating opportunities
for both basic research and applications. The coherent destruction of
tunneling \cite{Grossmann1991} and quantum charge pumping
\cite{Thouless1983,Brouwer1998,Altshuler1999}, \textit{i.e.} the
generation of a dc current at zero bias voltage due to quantum
interference \cite{Buttiker2006}, are prominent examples of the wealth
of phenomena driving this rapidly advancing area of research. By means
of time-dependent gatings  \cite{Switkes1999,Blumenthal2007}, surface
acoustic waves \cite{Leek2005} or illumination with a laser field
\cite{Srivastava2004}, experiments have probed different facets of ac
transport and light-matter interaction. On the other hand, theoretical
insights keep opening exciting directions, from interaction-induced quantum
pumping \cite{Reckermann2010} to everlasting oscillations
\cite{Kurth2010}. 

Among many other benefits, the advent of graphene \cite{CastroNeto2009}
and carbon nanotubes \cite{SaitoBook} provide an outstanding arena for the
study of ac transport in highly coherent, low dimensional
systems. Adiabatic quantum pumps \cite{Prada2009,Zhu2009}, ac controlled
Fabry-P\'{e}rot resonators \cite{Rocha2010} and photodetectors
\cite{Xia2009} are among the new breed of carbon-based
devices. Furthermore, a few remarkable studies on Dirac fermions
interacting with linearly \cite{Syzranov2008,Naumis2008} and circularly
\cite{Oka2009,Abergel2009} polarized monochromatic light pointed out
striking non-perturbative non-adiabatic effects: a laser field could
lead to the opening of dynamical gaps \cite{Syzranov2008} in
graphene. For circularly polarized light, a further gap was shown to
develop at the Dirac point \cite{Oka2009}.  
Many open questions remain: which experimental setup would unveil these
phenomena and how could we tune them?

In this Letter, we aim at ellucidating these questions by analyzing the
interaction between electrons and a monochromatic laser field of
\textit{arbitrary} polarization in graphene. Using a Floquet approach
one finds that a laser of frequency $\Omega$ induces the lifting of
degeneracies between (Floquet) states of the combined electron-photon
system.
Here, a careful analysis allows for the tuning of a feasible parameter
range (laser frequency, power and polarization). Our predictions show
that these effects are within the reach of mid-infrared laser technology
in a transport setup, thereby opening promising prospects for
graphene-based opto-electronic devices.

In graphene, the low energy electronic states contributing to transport
are close to the Dirac points $\mathbf{K}$ and
$\mathbf{K}^{\prime}$. Since we consider a clean sample and given that
the ac field does not introduce any inter-valley coupling, we can
describe both points separately. In the $\mathbf{K}$-valley, those
states can be accurately described by the $\mathbf{k\cdot p}$
approximation through the envelope wave-function
$\Psi=(\Psi_{A},\Psi_{B})^{\mathbf{T}}$, where the two components refer
to the interpenetrating sublattices $A$ and $B$ \cite{kp}. The time
periodic electromagnetic field, with period $T=2\pi /\Omega$, is
included as a monochromatic plane wave traveling along the $z$-axis,
perpendicular to the plane defined by the graphene sheet. The magnetic
vector potential is thus written as $\mathbf{A}(t)=
\mathrm{Re}\left\{\mathbf{A}_0e^{-\mathrm{i}\Omega t}\right\}$, where
$\mathbf{A}_0=A_0(1,e^{\mathrm{i}\varphi})$ refers to the intensity and
polarization $\varphi$ of the field. For this choice of parameters,
$\varphi=0$ yields a linearly polarized field $\mathbf{A}(t)= A_0\cos
\Omega t~(\mathbf{x}+\mathbf{y})$ while $\varphi=\pi /2$ results in a
circularly polarized field $\mathbf{A}(t)=A_0(\cos \Omega
t~\mathbf{x}+\sin \Omega t~\mathbf{y})$. Therefore, the graphene
electronic states in the presence of the ac field are encoded in the
Hamiltonian:

\begin{equation}
\mathcal{H}(t)=v_F\mathbf{\hat{\sigma}}\cdot\left[\mathbf{p}-
e\mathbf{A}(t)\right]. 
\label{eq_Hamiltonian} 
\end{equation}

Here $v_{F}\simeq 10^{6}\unit{m/s}$ denotes the Fermi velocity and
$\mathbf{\hat{\sigma}}=(\hat{\sigma}_x,\hat{\sigma}_y)$ the Pauli
matrices describing the pseudospin degree of
freedom. The suitable wavelength of the laser field lies
within the mid-infrared region, \textit{i.e.} $\lambda \simeq 9\unit{\mu
m}$, $\hbar\Omega\simeq 140 \unit{meV}$ and we choose two intensity
values $I\simeq 32 \unit{mW/\mu m^2}$ and $I\simeq 130 \unit{mW/\mu
m^2}$ such that $eA_{0}v_{F}/\hbar\Omega \simeq 0.17$ and $0.33$
respectively.

As will be made clear later on, a correct description of our problem
crucially requires a solution valid beyond the adiabatic
approximation. Hence, the Floquet theory \cite{Kohler2005}
represents an appropriate approach to describe such electron-photon
scattering processes. The resulting picture is that of an effective
time-independent Hamiltonian in a higher-dimensional space, the
so-called Floquet space, defined as the direct product between
the usual Hilbert space and the space of $2\pi/\Omega$-time periodic functions.  
This space is spanned by the states $\left\{\left\vert\mathbf{k},
n\right\rangle_{\pm}\right\}$, where $\mathbf{k}$ is the electronic wave  
vector, $n$ is the Fourier index, and the subindex refers to the
alignment of the pseudo-spin with respect to the momentum. In this
basis, the problem to solve is identical to the time-independent 
Schr\"{o}dinger equation with Floquet Hamiltonian $H_F=H-\mathrm{i}\hbar
\partial_t$. The power of recursive Green's function techniques
\cite{PastawskiRMF2001} can be exploited to obtain both the dc component
of the conductance and density of states from the so-called Floquet
Green's functions \cite{FoaTorres2005}.


Previous studies \cite{Syzranov2008,Oka2009} have considered lasers
either in the far infra-red ($\hbar\Omega=29\unit{meV}$)
\cite{Syzranov2008} or in the visible range \cite{Oka2009}. In the first
case, the predicted gaps where of about $6\unit{meV}$ and the authors
considered the photocurrents generated in a p-n junction. On the other
hand, for a laser in the visible range, the photon energy of about
$2\unit{eV}$ is much larger than the typical optical phonon energies 
($170\unit{meV}$), severe corrections to the transport properties due to dissipation
of the excess energy via electron-(optical)phonon interactions are expected. Besides,
appreciable effects in this last case required a power above $1 \unit{W/\mu m^2}$, which could
compromise the material stability. To overcome both limitations we
quantitatively explore the interaction with a laser in the mid-infrared
range ($\lambda=5 - 10 \unit{\mu m}$), where photon energies can
be made smaller than the typical optical phonon energy while keeping a
much lower laser power. Furthermore, we show that the polarization,
whose role was overlooked, can be used as a control variable.

\begin{figure}[tbp]
\includegraphics[width=8.6cm]{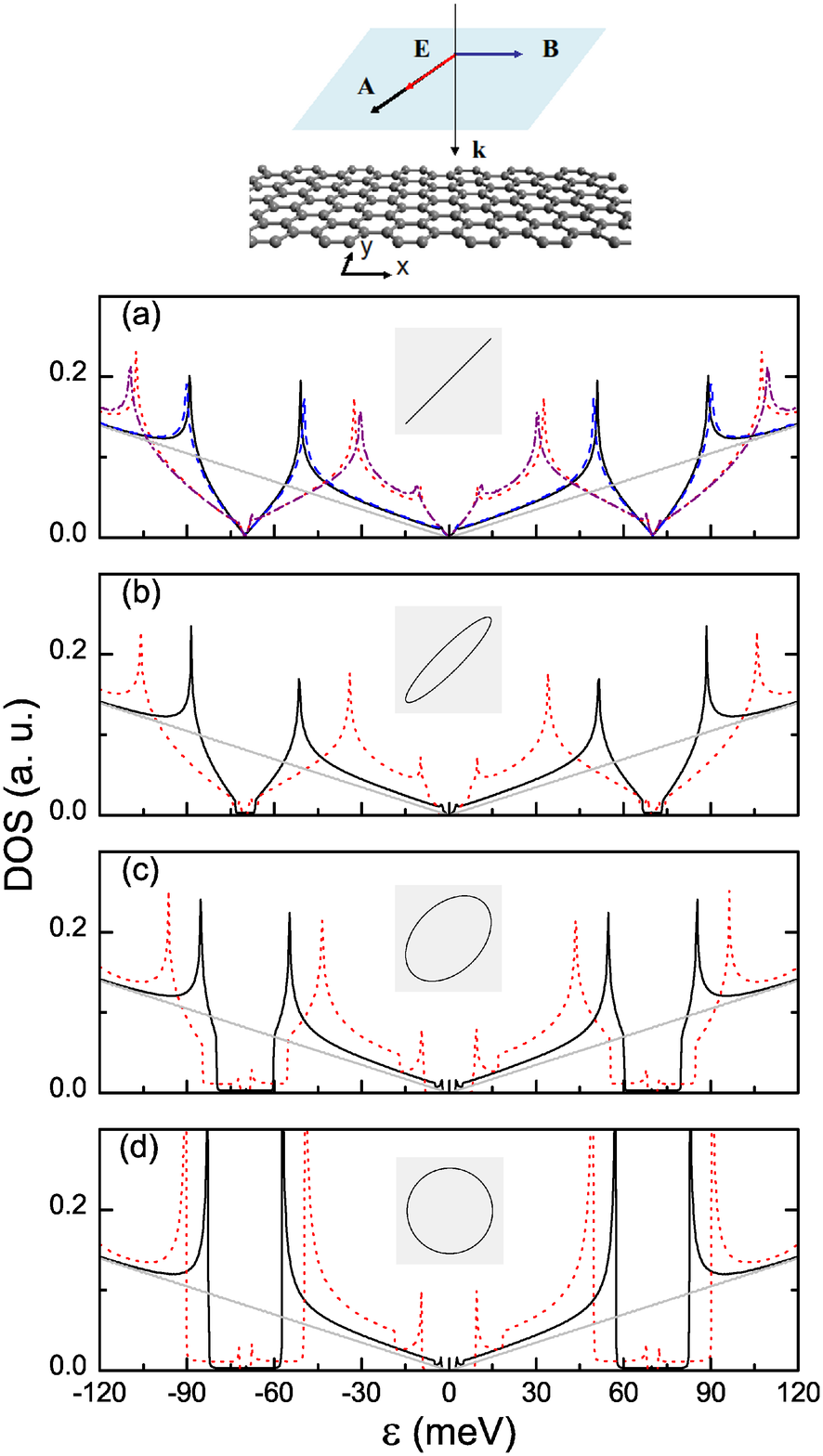}
\caption{(Color online) Scheme of the considered setup where a laser
 field with $\hbar\Omega=140\unit{meV}$ is applied perpendicular to a
 graphene mono-layer. Density of states for (a) linear, (b)
 $\varphi =0.125\pi$, (c) $\varphi =0.375\pi$, and (d) circular
 polarizations. The black solid line is for $I=32 \unit{mW/\mu m^2}$
 while the red dashed line corresponds to $I=130 \unit{mW/\mu m^2}$. For
 linear polarization these results are compared with those of a
 tight-binding calculation for a system with $5\times 10^4$ channels
 (blue and purple lines). For reference the zero-field DOS is shown in
 solid grey.}
\label{fig-1}
\end{figure}

Figure \ref{fig-1} shows how the dc density of states (DOS) 
evolves as the polarization changes. Already for
linear polarization (upper panel) one can find some surprises not
reported before: i) close to the Dirac point, the DOS is enhanced as
compared to the case without laser (red dotted line). The DOS increases
linearly but with a different slope. ii) Close to  $\pm\hbar\Omega /2$,
stronger effects lead to a depletion of the states but without reaching
a full gap. The counterpart of this depletion is given by the peaks in
the DOS on each side of the depletion area. More accurate calculations
based on a tight-binding model in a $\unit{\mu m}$ sized sample
interacting with a linelarly polarized field confirm these findings (top
panel).

The successive panels in Fig. \ref{fig-1} show how these features change
with the laser polarization: i) close to the Dirac point a gap opens
and the structure of the side peaks is severely modified. ii) The
depletion areas around $\pm\hbar\Omega /2$ transform into gaps whose
widths are maximum for circular polarization. A closer look at
these figures show that the mentioned gaps are areas with a reduced
but non-vanishing DOS. 

\begin{figure}[tbp]
\includegraphics[width=8.0cm]{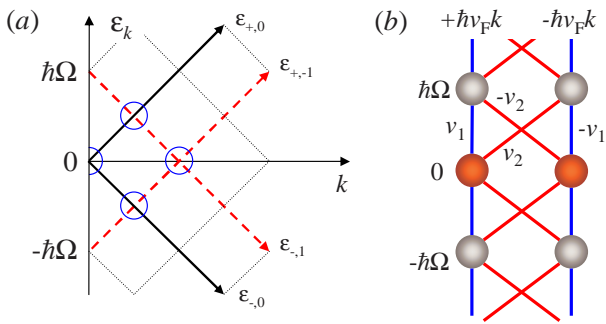}
\caption{(Color online) (a) Scheme of the quasienergies as a function of
 $k$. The relevant crossing regions, marked by blue and red circles,
 yield the dynamical and central gap respectively. (b) Representation of
 the Floquet Hamiltonian for the $\mathbf{k\cdot p}$ approach. Here, the
 direct hopping $v_1=\frac{eA_0v_F}{2}(\cos\alpha+e^{-\mathrm{i}\varphi}
 \sin\alpha)$ sets the transition
 amplitude between Floquet states conserving pseudospin. In contrast, 
the off-diagonal term $v_2=\mathrm{i}\frac{eA_0v_F}{2}
 (e^{-\mathrm{i}\varphi}\cos\alpha -\sin\alpha)$ introduces an inelastic
 back-scattering process that simultaneously enables both $m$ and
 pseudospin transitions.} 
\label{fig-2}
\end{figure}

To rationalize the behavior observed in Fig. \ref{fig-1} we resort to
the Floquet picture explained above. A scheme with the relevant states
close to the Dirac point is shown in Fig. \ref{fig-2}. On the left panel, one
can see the dispersion relation for the states $\left\{\left\vert
\mathbf{k},n\right\rangle_{\pm}\right\}$ for $n=0$ (black solid line)
and $n=\pm 1$ (red line), the scheme on the right panel shows the states
represented by circles and their corresponding interactions with full
lines. The effects of the ac field are expected to be stronger at the
crossing points between these lines (whenever they have a non vanishing
Hamiltonian matrix element), leading to the opening of energy gaps at
those points. From a geometrical argument one can see that the crossing of 
the states differing in one photon lies exactly at $\pm\hbar\Omega /2$. 
These degeneracies are lifted by the ac field leading to the gaps observed in Fig. \ref{fig-1}. 
This mechanism is reminiscent of an inelastic Bragg reflection.
An estimation of the resulting gap gives: 

\begin{equation}
\Delta_{k=\Omega /2v_F}\simeq eA_0v_F\sqrt{1-\cos(\varphi)
\sin(2\alpha)},
\label{eq-dynamical}  
\end{equation}
where $\alpha=\tan ^{-1}(k_y/k_x)$. Equation \ref{eq-dynamical} shows that the effect is
linear in the field strength. Interestingly, after summing up over all
the directions in the 2d $\mathbf{k}$-space, we see that no net gap opens
in the linearly polarized case since in the orientation $\alpha=\pi /4$
both states become degenerate. However, as can be seen in
Fig. \ref{fig-1}, there is a strong modification in the DOS that
resembles the usual Dirac point for a Fermi energy around $\hbar\Omega
/2$. Changing the polarization away from the linear case, one sees that  a \textit{dynamical gap}
\cite{Syzranov2008} opens and reaches its maximum $\Delta
_{\mathrm{max}}\simeq 23 \unit{meV}$ ($46 \unit{meV}$) for $I=32
\unit{mW/\mu m^2}$ ($130 \unit{mW/\mu m^2}$) in the circularly polarized
case. 

Now, let us account for the phenomenon occuring at the Dirac point. 
Oka and Aoki \cite{Oka2009} predicted that
circularly polarized light would induce a further gap around the Dirac
point, in agreement with our Fig. \ref{fig-1}-d. Here we see that these
strong modifications extend all the way up to linear polarization. A
careful analysis shows that it is a higher-order effect. The leading
contribution comes from the states connecting the degenerate ones at the
Dirac point (there are four of these paths as can be seen by analyzing
Fig. \ref{fig-2}-b). The calculated gap reads:

\begin{equation}
\Delta _{k=0}\simeq \frac{8}{\hbar \Omega }\mathrm{Re}\left\{ v_{1}v_{2}^{\ast }\right\} =2\frac{(eA_{0}v_{F})^{2}}{\hbar \Omega }%
\sin \varphi,
\label{eq-central}
\end{equation}
which gives a quadratic dependence with the field strength and is
inversely proportional to the frequency. Note also that there is no
dependence of the gap with the orientation of the
$\mathbf{k}$-vector. The polarization yields a maximum gap for the
circularly polarized laser field while no net gap opens for the linearly
polarized case. However in Fig. \ref{fig-1}-a noticeable 
corrections around the Dirac point are observed even for linear
polarization. A deeper analysis shows that this is due to the
field-induced lifting of the degeneracies between the states $\left\vert
\mathbf{k},\pm 1 \right\rangle_{\pm}$ around $\varepsilon =0$. This
effect is of the same order as the correction given by
Eq. \ref{eq-central} and becomes stronger as the photon energy
is reduced.  

\begin{figure}[tbp]
\includegraphics[width=8.0cm]{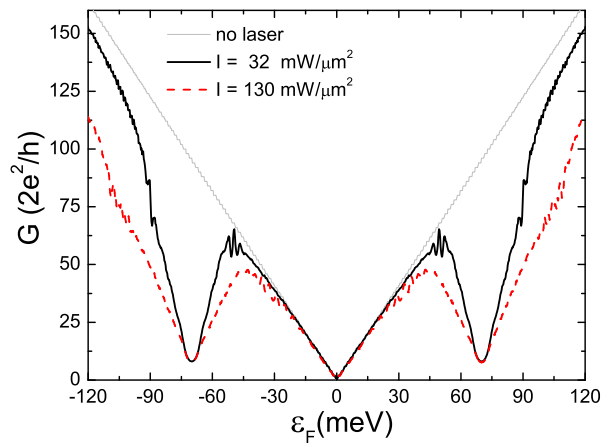}
\caption{(Color online) dc conductance through a graphene stripe of
 $1\mu m$ times  $1\mu m$ in the presence of a linearly polarized laser
 as a function of the Fermi energy. The black solid line is for $I=32
 \unit{mW/\mu m^2}$ while the red dashed line corresponds to $I=130
 \unit{mW/\mu m^2}$.}
\label{fig-3}
\end{figure}

Up to now, we have shown how the laser field modifies the electronic
structure of 2d graphene that would be observed in any experiment
carried out over a time much larger than the period $T$. A relevant
question is if these effects would be observable in a transport
experiment. To such end we compute the transport response at zero temperature using Floquet
theory \cite{FoaTorres2005} applied to a $\pi$ orbitals Hamiltonian. The
electromagnetic field is accounted through the Peierls' substitution which 
introduces an additional phase in the hopping
$\gamma_{ij}$ connecting adjacent sites $\mathbf{r}_i$ and
$\mathbf{r}_j$:  $\gamma_{ij}=
\gamma_{0}\exp\left(\mathrm{i}\frac{2\pi}{\Phi_0}\int_{
\mathbf{r}_i}^{\mathbf{r}_j}\mathbf{A}(t)\cdot\mathrm{d}\mathbf{r}\right)$,  
where $\gamma_0\simeq 2.7\unit{eV}$ is the hopping amplitude at zero
field and $\Phi_0$ is the quantum of magnetic flux. For computational 
convenience we use an armchair edge structure with the vector potential
$\mathbf{A}(t)=A_x\cos\Omega t~\mathbf{x}+A_y\sin\Omega t~\mathbf{y}$.

Calculating the two-terminal dc component of the conductance as a
function of Fermi energy is not, \textit{a priori}, computationally easy. 
An efficient solution is achieved by decomposing the  Hamiltonian
into independent transversal channels as explained in \cite{Rocha2010}. 
Although this decomposition is preserved by the interaction with the laser only
for linear polarization ($A_y=0$), it is already enough to give a flavor on the
transport effects. Figure \ref{fig-3} gives the computed dc conductance
\cite{Kohler2005,FoaTorres2005} for a stripe $1\unit{\mu m}$ wide and
$1\unit{\mu m}$ long. One can appreciate in Fig. \ref{fig-3} that the 
strong depletion areas observed in the DOS of Fig. \ref{fig-1} are
indeed mirrored in the dc component of the conductance. Accordingly, 
the predicted dynamical gaps can be unveiled through transport measurements.

In summary, we have shown that it is possible to
use a laser field in the mid-infrared range to tune the electronic
structure of graphene and its electrical response. 
The modifications are predicted to arise both around the Dirac point and
at $\pm \hbar \Omega /2$. Moreover, since the results are
strongly dependent on the laser polarization, it may be used as a control
parameter. We encourage  experimentalists to pursue this exciting line
of research.

We acknowledge correspondence with J. Kono, discussions with
G. Usaj and support by SeCyT-UNC, ANPCyT-FonCyT . LEFFT
acknowledges 
the support from the Alexander von Humboldt Foundation. 



\end{document}